\magnification=1200
\def\d{\delta}
\def\g{\gamma}
\def\k{\kappa}\def\o{\omega}\def
\p{\pi}\def\t{\tau}
\def\x{\xi}

\def\G{\Gamma}

\def\id{\equiv}\def\mo{{-1}}\def\ha{{1\over 2}}

\def\coo{coordinates }

\def\pb{Poisson brackets }

\def\poi{Poincar\'e }

\def\tl{transformation law }
\def\wrt{with respect to }

\def\section#1{\bigskip\noindent{\bf#1}\smallskip}
\def\nota{\footnote{$^\dagger$}}

\def\PL#1{Phys.\ Lett.\ {\bf#1}}
\def\PRL#1{Phys.\ Rev.\ Lett.\ {\bf#1}}
\def\PR#1{Phys.\ Rev.\ {\bf#1}}\def\CQG#1{Class.\ Quantum Grav.\ {\bf#1}}

\def\AoP#1{Ann.\ Phys.\ {\bf#1}}
\def\grq#1{{\tt gr-qc/#1}}\def\hep#1{{\tt hep-th/#1}}

\def\ref#1{\medskip\everypar={\hangindent 2\parindent}#1}
\def\beginref{\begingroup
\bigskip
\centerline{\bf References}
\nobreak\noindent}

\def\bv{{\bf v}}
\def\DLT{deformed Lorentz transformations }

%%%%%%%%%%%%%%%%%%%%%%%%%%%%%%%%%%%%%%%%%%%%%%%%%%%%%%%%%%%%%%%%%%
{\nopagenumbers
\line{}
\vskip80pt
\centerline{\bf A spacetime realization of $\k$-\poi algebra}
\vskip60pt
\centerline{{\bf S. Mignemi}\nota{e-mail: smignemi@unica.it}}
\vskip10pt
\centerline {Dipartimento di Matematica, Universit\`a di Cagliari}
\centerline{viale Merello 92, 09123 Cagliari, Italy}
\smallskip
\centerline{and INFN, Sezione di Cagliari}

\vskip100pt
\centerline{\bf Abstract}
\vskip10pt
{\noindent
We study a Hamiltonian realization of the phase space of $\k$-\poi
algebra that yields a definition of velocity consistent with the
deformed Lorentz symmetry.
We are also able to determine the laws of transformation of
spacetime coordinates and to define an invariant spacetime metric,
and discuss some possible experimental consequences.}
\vskip100pt
%{\noindent P.A.C.S. Numbers: 04.60.-m 04.70.Bw 11.15.-q}
\vfil\eject}
Doubly special relativity (DSR) is a class of models which aim to give an
effective description of quantum gravity effects on particle kinematics at
energies near the Planck scale, $\k\sim10^{19}$ GeV, by postulating a nonlinear
(deformed) action of the Lorentz group on momentum space [1].
In spite of the recent advances
in the understanding of this proposal, some problems are still open, as for
example what should be the realization of the theory in position space.
Related to this is the problem of defining the velocity and the dynamics
of a particle in a way consistent with the deformed Lorentz transformations.

In a recent paper [2], we have proposed a method for introducing a Hamiltonian
structure for DSR models in such a way that the velocity of a particle,
defined classically in terms of the proper-time derivatives of the coordinates
as $\bv=\dot x_i/\dot x_0$, coincides with the definition proposed in ref.\ [3],
based on the observation that the velocity can be viewed
as the parameter of the (deformed) boosts.
The formalism of ref.\ [2] also induces in a natural way a realization of
the deformed Lorentz symmetry in position space.

Our prescription worked well for the model of ref.\ [4], but led to
inconsistencies in the case of the $\k$-\poi model of ref.\ [5].
In particular, the expression for the velocity derived in [2] did not
transform in the correct way and consequently
it was not possible to define an invariant line element.
In [6] it was noticed that this problem can be solved in general  by imposing
further constraints on the symplectic structure.

In this note we wish to show how the results of [6] can be
applied also to the case of the $\k$-\poi model. We refer to [2] and [6] for
further motivations and technical details.
For simplicity, we work in $1+1$ dimensions. We denote
the position and the momentum of a particle as $q_a$ and $p_a$, $a=0,1$.
\bigbreak

The $\k$-\poi model [5] is defined by the following nonlinear \tl for the
momentum of a particle under a boost of parameter $\x=\tanh v$:
$$p'_0=p_0+\k\log\G,\qquad p'_1={p_1\cosh\x+{\k\over2}\left(1-e^{-2p_0/\k}+
{p_1^2\over\k^2}\right)\sinh\x\over\G}\, ,\eqno(1)$$
where
$$\G=\ha\left(1+e^{-2p_0/\k}-{p_1^2\over\k^2}\right)+\ha\left(1-e^{-2p_0/\k}
+{p_1^2\over\k^2}\right)\cosh\x+{p_1\over\k}\sinh\x.\eqno(2)$$
In infinitesimal form the \tl (1) reads
$$\d p_0=p_1,\qquad \d p_1={\k\over2}\left(1-e^{-2p_0/\k}-
{p_1^2\over\k^2}\right).\eqno(3)$$

The Hamiltonian for a free particle can be identified with the Casimir
invariant
$$
H={m^2\over2}={e^{2p_0/\k}\over2}\left[{\k^2\over4}\left(1-e^{-2p_0/\k}+
{p_1^2\over\k^2}\right)^2-p_1^2\right].\eqno(4)
$$
The physical mass $M$ of the particle, i.e.\ its energy at rest, is related
to $m$ by $m=\k\sinh(M/\k)$.

The Hamiltonian is not uniquely defined. For example, in ref.\ [7] it was
chosen as
$$\tilde H={\k\over2}e^{p_0\k}\left(1+e^{-2p_0/\k}-{p_1^2\over\k^2}\right),
\eqno(5)$$
which is related to ours by $H=(\tilde H^2-\k^2)/2$.

In DSR models, the momentum $p_a$ can be related by a nonlinear transformation
to an unphysical momentum $\pi_a$ that transforms linearly under \DLT [8]. In
our case,
$$\p_0={\k\over2}\,e^{p_0/\k}\left(1-e^{-2p_0/\k}+{p_1^2\over\k^2}\right),
\qquad\p_1=e^{p_0/\k}p_1.\eqno(6)$$
According to ref.\ [3], the definition of the velocity $\bv$ compatible with
its role of parameter of the Lorentz transformations is then
$$\bv={\p_1\over\p_0}={2p_1/\k\over1-e^{-2p_0/\k}+{p_1^2/\k^2}}.\eqno(7)$$
This expression for the velocity can be obtained from the basic definition
$v=\dot q_1/\dot q_0$, where $\dot q_a$ are the derivatives
of the coordinates \wrt the proper time derived from the Hamilton equations,
if one postulates the standard $\k$-\poi symplectic structure [7,2]
$$\o_{00}=1,\quad\o_{01}=-{p_1\over\k},\quad\o_{10}=0,\quad\o_{11}=-1,
\eqno(8)$$
with $\o_{ab}\id\{q_a,p_b\}$.
However, one may choose different \pb leading to the same
expression for the velocity. In particular, in [6] it was shown that in order
for the velocity to transform in the correct way under deformed boosts, one
must impose
some further conditions on the $\o_{ab}$. When these conditions hold, it is
also possible to define a (momentum-dependent) metric invariant under deformed
Lorentz transformations.
In the present case, the conditions of [6] are not satisfied by (8),
but rather by
$$\o_{00}=\ha(1 + e^{-2p_0/\k} + {p_1^2\over\k^2}),\quad\o_{01}=
-{p_1\over\k}e^{-2p_0/\k},\quad\o_{10}={p_1\over\k},\quad\o_{11}=
-e^{-2p_0/\k}.\eqno(9)$$
Given the symplectic structure (9),
the Jacobi identities imply that the \coo obey nontrivial \pb
$$\{q_0,q_1\}=2{p_1q_0\over\k^2}-\left(1+{p_1^2\over\k^2}\right)
{q_1\over\k},\eqno(10)$$
and transform as
$$\d q_0=q_1-{p_1\over\k}q_0,\quad\d q_1=q_0-{p_1\over\k}q_1.\eqno(11)$$

Moreover, the Hamilton equations arising from (4) and (8) read
$$\eqalignno{
\dot q_0&={\k\,e^{2p_0/\k}\over8}\left(1+e^{-2p_0}-
{p_1^2\over\k^2}\right)^2\left(1-e^{-2p_0/\k}+{p_1^2\over\k^2}\right),\cr
\dot q_1&={e^{2p_0/\k}\over4}\left(1+e^{-2p_0}-
{p_1^2\over\k^2}\right)^2p_1,&(12)}$$
from which one can recover the velocity (7).
It is also easy to see that the line element
$$
d\t^2={4\,e^{-2p_0/\k}\over(1+e^{-2p_0}-p_1^2/\k^2)^4}(dq_0^2-dq_1^2)
\eqno(13)$$
is invariant under the infinitesimal \DLT (3) and (11). Contrary to other
known cases [2,6], the metric (13) depends on both components of $p_a$, and not
on the energy only. However, comparing with (5), one may write (13) in the
simpler form
$$ds^2={1\over4}\left(1+{2m^2\over\k^2}\right)^{-2}\ e^{2p_0/\k}(dq_0^2-
dq_1^2).\eqno(14)$$

The transformations (10) that, combined with (1), leave (12) invariant can also
be written in finite form as
$$q'_0={q_0\cosh\x+q_1\sinh\x\over\G},\qquad q'_1={q_1\cosh\x+q_0\sinh\x\over\G}.
\eqno(15)$$
Hence the coordinate transformations take the form of a product of standard
Lorentz transformations with a function of the momentum.

The transformations (15) imply a modification of the relativistic formula for
time dilation. For example, it is easy to see that the relation between the time
$T$ measured in the laboratory and the time $T_0$ measured in the rest frame of
a particle is given by
$$T={\g\over\G_0}\ T_0,\eqno (16)$$
where $\g=\cosh\x=(1-{\bf v}^2)^{-1/2}$ and
$$\G_0\id\G(p_1=0)=\ha(1+e^{-2M/\k})+{\g\over2}(1-e^{-2M/\k}).\eqno (17)$$
Hence $T$ becomes a function both of the velocity and the momentum, or
equivalently the mass, of the
particle, giving rise to corrections of order $p_0/\k$ to the measured lifetime
of high-velocity particles with respect to the relativistic formula, which in
principle may be susceptible of experimental verification.
In fact, for $\g\gg1$, one has in first approximation in $\k^\mo$,
$T\sim\g\left(1-{M\g\over\k}\right)T_0\sim\g\left(1-{p_0\over\k}\right)T_0$.
In this approximation, the size of the corrections is given by the ratio of the
energy of the particle and the Planck energy.
If one takes for $\k$ the standard value of the Planck energy, $10^{19}$ GeV,
this would be an extremely small correction, not detectable experimentally
even for particles of energy of order 10 GeV.
However, in the context of some higher-dimensional theories, the effective
four-dimensional Planck energy $\bar\k$ could be lowered up to $10^3$ GeV [9], and
in this case corrections could be observed.

To our knowledge, the relativistic formula for time delay has been checked for
pions with $\g\sim2.4$, with a confidence of 0.4\% [10].
This fixes a lower limit for $\bar\k$ to $100$ GeV.
Improving the experimental limits could give evidence for a modification of
the time delay formula, only if $\bar\k$ is not much greater than this value.

We conclude by remarking that the line element (13) may also take the role of the
metric in a formulation of a $\k$-\poi extension of general relativity on the
lines of the gravity rainbow formalism of ref.\ [11].
In the present case, the metric would depend
not only on the energy, but also on the space component of the momentum (or
equivalently on the mass) of the particle. It  must be also pointed out that in the
present framework the speed of
light is independent of the energy, contrary to the case of ref.\ [11].

\section{Acknowledgments}
I wish to thank Maurizio Melis for useful discussions and the referee for pointing
out some errors in a first version of the paper.
\bigbreak
\beginref
\ref [1] For a recent review, see J. Kowalski-Glikman, \hep{0405273};
G. Amelino-Camelia, \grq{0412136}.

\ref [2] S. Mignemi, \PR{D68}, 065029 (2003).

\ref [3] P. Kosi\'nski and P. Ma\'slanka, \PR{D68}, 067702 (2003).

\ref [4] J. Magueijo and L. Smolin, \PRL{88}, 190403 (2002).

\ref [5] J. Lukierski, A. Nowicki, H. Ruegg and V.N. Tolstoy, \PL{B264},
331 (1991); J. Lukierski, A. Nowicki and H. Ruegg, \PL{B293}, 344
(1992); J. Lukierski, H. Ruegg and W.J. Zakrzewski, \AoP{243}, 90 (1995).

\ref [6] S. Mignemi, \grq{0403038}.

\ref [7] J. Lukierski and A. Nowicki, Acta Phys. Polon. {\bf B33}, 2537 (2002).

\ref [8] S. Judes and  M. Visser, \PR{D68}, 045001 (2003).

\ref [9] L. Randall and R. Sundrum, \PRL{83}, 3370 (1999).

\ref [10] A.J. Greenberg {\it et al.}, \PRL{23}, 1267 (1969).

\ref [11] J. Magueijo and L. Smolin, \CQG{21}, 1725 (2004).

\end